\documentclass[12pt,leqno]{article}
\usepackage{amssymb,amsmath,amsthm,graphicx}

\def\beq{\begin{equation}}
\def\eeq{\end{equation}}

\def\bTheta{{\boldsymbol{\Theta}}}

\def\bx{\mathbf{x}}

\def\b0{\mathbf{0}}
\def\bI{\mathbf{I}}
\def\bA{\mathbf{A}}

\def\bD{\mathbf{D}}
\def\bB{\mathbf{B}}
\def\bE{\mathbf{E}}
\def\be{\mathbf{e}}
\def\bF{\mathbf{F}}
\def\bG{\mathbf{G}}
\def\bH{\mathbf{H}}
\def\bJ{\mathbf{J}}
\def\bM{\mathbf{M}}
\def\bK{\mathbf{K}}
\def\bN{\mathbf{N}}
\def\bL{\mathbf{L}}
\def\bP{\mathbf{P}}
\def\bS{\mathbf{S}}
\def\bT{\mathbf{T}}
\def\bQ{\mathbf{Q}}
\def\bU{\mathbf{U}}
\def\bV{\mathbf{V}}
\def\bW{\mathbf{W}}
\def\bX{\mathbf{X}}
\def\bY{\mathbf{Y}}
\def\bZ{\mathbf{Z}}
\def\bdelta{{\boldsymbol{\delta}}}
\def\beps{{\boldsymbol{\varepsilon}}}
\def\bSigma{{\boldsymbol{\Sigma}}}
\def\tbX{\tilde{\mathbf{X}}}
\def\tbTheta{\tilde{\boldsymbol{\Theta}}}
\def\cF{\mathcal{F}}
\def\cG{\mathcal{G}}
\def\cO{\mathcal{O}}

\def\cR{\mathcal{R}}

\def\IE{{\mathbb E}}
\def\ha{\hat{a}}
\def\hb{\hat{b}}
\def\hR{\hat{R}}
\def\eps{\varepsilon}
\def\tA{\tilde{A}}
\def\tB{\tilde{B}}
\def\tC{\tilde{C}}
\def\tD{\tilde{D}}
\def\tR{\tilde{R}}
\def\tbA{\tilde{\mathbf{A}}}
\def\tbM{\tilde{\mathbf{M}}}
\def\tbN{\tilde{\mathbf{N}}}

\def\tbZ{\tilde{\mathbf{Z}}}
\def\tbT{\tilde{\mathbf{T}}}
\def\tbU{\tilde{\mathbf{U}}}
\def\tbV{\tilde{\mathbf{V}}}

\def\tbx{\tilde{\mathbf{x}}}
\def\ta{\tilde{a}}
\def\tb{\tilde{b}}
\def\tR{\tilde{R}}
\def\tu{\tilde{u}}
\def\tv{\tilde{v}}
\def\tx{\tilde{x}}
\def\ty{\tilde{y}}
\def\tz{\tilde{z}}
\def\ttheta{\tilde{\theta}}
\def\tbJ{\tilde{\mathbf{J}}}

\begin{document}

\title{Error analysis for circle fitting algorithms}

\author{A. Al-Sharadqah$^{1}$ and N. Chernov$^{1}$}

\footnotetext[1]{Department of Mathematics, University of Alabama at
Birmingham, Birmingham, AL 35294; chernov@math.uab.edu;
alsha1aa@uab.edu}

\date{}

\maketitle

\begin{abstract}
We study the problem of fitting circles (or circular arcs) to data
points observed with errors in both variables. A detailed error
analysis for all popular circle fitting methods -- geometric fit,
K{\aa}sa fit, Pratt fit, and Taubin fit -- is presented. Our error
analysis goes deeper than the traditional expansion to the leading
order. We obtain higher order terms, which show exactly why and by
how much circle fits differ from each other. Our analysis allows us
to construct a new algebraic (non-iterative) circle fitting
algorithm that outperforms all the existing methods, including the
(previously regarded as unbeatable) geometric fit.
\end{abstract}

\begin{center}
\end{center}

\section{Introduction}\label{secI}

Fitting circles and circular arcs to observed points is one of the
basic tasks in pattern recognition and computer vision, nuclear
physics, and other areas \cite{Be89, CL1, De72, Ka76, La87, Pr87,
Sp96, Ta91}. Many algorithms have been developed that fit circles to
data. Some minimize the geometric distances from the circle to the
data points (we call them geometric fits). Others minimize various
approximate (or `algebraic') distances, they are called algebraic
fits. We overview most popular algorithms in
Sections~\ref{SecGCF}--\ref{SecACF}.

Geometric fit is commonly regarded as the most accurate, but it can
only be implemented by iterative schemes that are computationally
intensive and subject to occasional divergence. Algebraic fits are
faster but presumably less precise. At the same time the assessments
on their accuracy are solely based on practical experience, no one
has performed a detailed theoretical comparison of the accuracy of
various circle fits. It was shown in \cite{CL2} that all the circle
fits have the same covariance matrix, to the leading order, in the
small-noise limit. Thus the differences between various fits can
only be revealed by a higher-order error analysis.

The purpose of this paper is to do just that. We employ higher-order
error analysis (a similar analysis was used by Kanatani \cite{Ka08a}
in the context of more general quadratic models) and show exactly
why and by how much the geometric circle fit outperforms the
algebraic circle fits in accuracy; we also compare the precision of
different algebraic fits. Section~\ref{SecEAGS} presents our error
analysis in a general form, which can be readily applied to other
curve fitting problems.

Finally, our analysis allows us to develop a new algebraic fit whose
accuracy exceeds that of the geometric fit. Its superiority is
demonstrated by numerical experiments.

\section{Statistical model}\label{SecSM}

We adopt a standard \emph{functional model} in which data points
$(x_1,y_1), \ldots, (x_n, y_n)$ are noisy observations of some
\emph{true points} $(\tx_1, \ty_1), \ldots, (\tx_n,\ty_n)$, i.e.
\beq \label{errors}
  x_i = \tx_i + \delta_i, \qquad
   y_i = \ty_i + \eps_i,
   \qquad i=1,\ldots,n,
  \eeq
where $(\delta_i, \eps_i)$ represent isotropic Gaussian noise.
Precisely, $\delta_i$'s and $\eps_i$'s are i.i.d.\ normal random
variables with mean zero and variance $\sigma^2$.

The true points $(\tx_i, \ty_i)$ are supposed to lie on a
`true circle', i.e.\ satisfy
\beq
  (\tx_i - \ta)^2 + (\ty_i - \tb)^2 = \tR^2,
   \qquad i=1,\ldots,n,
\eeq
where $(\ta, \tb, \tR)$ denote the `true' (unknown)
parameters. Therefore
\beq
   \tx_i = \ta + \tR \cos\varphi_i,
   \qquad \ty_i = \tb + \tR \sin\varphi_i,
\eeq
where $\varphi_1, \ldots, \varphi_n$ specify the locations of the
true points on the true circle. The angles $\varphi_1, \ldots,
\varphi_n$ are regarded as fixed unknowns and treated as additional
parameters of the model (called incidental or latent parameters).
For brevity we denote
\beq \label{tuv}
    \tu_i=\cos\varphi_i=(\tx_i-\ta)/\tR, \qquad
    \tv_i=\sin\varphi_i=(\ty_i-\tb)/\tR .
\eeq
Note that $\tu_i^2 + \tv_i^2 =1$ for every $i$.

\medskip\emph{Remark}. In our paper $\delta_i$ and $\eps_i$  have common variance $\sigma^2$, i.e.\ our noise is homoscedastic. In many studies the noise is heteroscedastic \cite{Meer04, YW04}, i.e.\ the normal vector $(\delta_i, \eps_i)$ has point-dependent covariance matrix $\sigma^2C_i$, where $C_i$ is known and depends on $i$, and $\sigma^2$ is an unknown factor. Our analysis can be extended to this case, too, but the resulting formulas will be somewhat more complex, so we leave it out.

\section{Geometric circle fits}\label{SecGCF}

A standard approach to fitting circles to 2D data is based on
orthogonal least squares, it is also called \emph{geometric fit}, or
\emph{orthogonal distance regression} (ODR). It minimizes the
function
\beq \label{cF}
       \cF(a,b,R) =  \sum d_i^2,
\eeq
where $d_i$ stands for the distance from $(x_i,y_i)$ to the circle,
i.e.
\beq \label{di}
       d_i= r_i-R, \qquad r_i=\sqrt{(x_i-a)^2+(y_i-b)^2},
\eeq
where $(a,b)$ denotes the center, and $R$ the radius of the circle.

In the context of the functional model, the geometric fit returns the
maximum likelihood estimates (MLE) of the circle parameters
\cite{Ch65}, i.e.
\beq  \label{argmin}
   (\ha_{\rm MLE}, \hb_{\rm MLE}, \hR_{\rm MLE}) =
   \, {\rm argmin}\, \cF(a,b,R).
\eeq

A major concern with the geometric fit is that the above
minimization problem has no closed form solution. All practical
algorithms of minimizing $\cF$ are iterative; some implement a
general Gauss-Newton \cite{Ch65, Jo94} or Levenberg-Marquardt
\cite{CL1} schemes, others use circle-specific methods proposed by
Landau \cite{La87} and Sp\"{a}th \cite{Sp96}. The performance of
iterative algorithms heavily depends on the choice of the initial
guess. They often take dozens or hundreds of iterations to converge,
and there is always a chance that they would be trapped in a local
minimum of $\cF$ or diverge entirely. These issues are explored in
\cite{CL1}.

A peculiar feature of the maximum likelihood estimates $(\ha, \hb,
\hR)$ of the circle parameters is that they have infinite moments \cite{C5},
i.e.\
\beq \label{inf}
   \IE(|\ha|) = \IE(|\hb|) = \IE(\hR) = \infty
\eeq
for any set of true values $(\ta, \tb, \tR)$; here $\IE$ denotes the mean value. This happens because the
distributions of these estimates have somewhat heavy tails, even
though those tails barely affect the practical performance of the
MLE (the same happens when one fits straight lines to data with
errors in both variables \cite{An76, AS82}).

To ensure the existence of moments one can adopt a different
parameter scheme. An elegant scheme was proposed by Pratt
\cite{Pr87} and others \cite{GGS94}, which describes circles by an
algebraic equation
\beq     \label{ABCDcircle}
      A(x^2+y^2) + Bx + Cy + D = 0
\eeq
with an obvious constraint $A\neq 0$ (otherwise this equation
describes a line) and a less obvious constraint $B^2+C^2-4AD>0$. The
necessity of the latter can be seen if one rewrites equation
\eqref{ABCDcircle} as
\beq
  \Bigl(x - \frac{B}{2A}\Bigr)^2+
  \Bigl(y - \frac{C}{2A}\Bigr)^2-
  \frac{B^2+C^2-4AD}{4A^2}=0.
\eeq
It is clear now that \eqref{ABCDcircle} defines a circle if and only
if $B^2+C^2-4AD>0$.

As the parameters $(A,B,C,D)$ only need to be determined up to a
scalar multiple, it is natural to impose a constraint
\beq  \label{constr}
        B^2+C^2-4AD = 1,
\eeq
because it automatically ensures $B^2+C^2-4AD>0$. The constraint
\eqref{constr} was first proposed by Pratt \cite{Pr87}. Under this
constraint, the parameters $A,B,C,D$ are essentially bounded, see
\cite{CL1}, and their maximum likelihood estimates can be shown to
have finite moments.

The equation \eqref{ABCDcircle}, under the constraint
\eqref{constr}, conveniently describes all circles and lines (the
latter are obtained when $A=0$); the inclusion of lines is necessary
to ensure the existence of the least squares solution \cite{CL1,
Ni02, ZC06a}.

After one estimates the algebraic circle parameters $A,B,C,D$, they
can be converted to the natural parameters via
\beq \label{conversion}
     a=-\frac{B}{2A},\qquad b=-\frac{C}{2A},
     \qquad R^2=\frac{B^2+C^2-4AD}{4A^2}.
\eeq

\section{Algebraic circle fits}\label{SecACF}

An alternative to the complicated geometric fit is made by fast
non-iterative procedures called \emph{algebraic fits}. We describe
three most popular algebraic circle fits below.

\medskip\noindent\textbf{K{\aa}sa fit}. One can find
a circle by minimizing the function
\begin{align} \label{Kasa1}
    \cF_{\rm K} &= \sum (r_i^2-R^2)^2\nonumber\\
    &=\sum (x_i^2+y_i^2 -2ax_i-2by_i +a^2+b^2-R^2)^2.
\end{align}
In other words, one minimizes $\cF_{\rm K} = \sum f_i^2$, where $f_i
= r_i^2 - R^2$ is the so called \emph{algebraic distance} from the
point $(x_i,y_i)$ to the circle. A change of parameters $B=-2a$,
$C=-2b$, $D=a^2+b^2-R^2$ transforms \eqref{Kasa1} to a linear least
squares problem minimizing
\beq \label{Kasa2}
    \cF_{\rm K} = \sum (z_i+Bx_i+Cy_i+D)^2,
\eeq
where we denote $z_i = x_i^2+y_i^2$ for brevity (we intentionally omit symbol $A$ here to make our formulas consistent with the subsequent ones). Now the problem
reduces to a system of linear equations (normal equations) with
respect to $B,C,D$ that can be easily solved, and then one recovers
the natural circle parameters $a,b,R$ via \eqref{conversion}.

This method was introduced in the 1970s by Delogne \cite{De72} and
K{\aa}sa \cite{Ka76}, and then rediscovered and published
independently by many authors, see references in \cite{CL1}. It
remains popular in practice. We call it \emph{K{\aa}sa fit}.

The K{\aa}sa method is perhaps the fastest circle fit, but its
accuracy suffers when one observes incomplete circular arcs
(partially occluded circles); then the K{\aa}sa fit is known to be
heavily biased toward small circles \cite{CL1}. The reason for the
bias is that the algebraic distances $f_i$ provide a poor
approximation to the geometric distances $d_i$; in fact,
\beq
  f_i = (r_i-R)(r_i+R)=d_i(2R+d_i) \approx 2Rd_i,
\eeq
hence the K{\aa}sa fit minimizes $\cF_{\rm K} \approx 2R^2\sum
d_i^2$, and it often favors smaller circles minimizing $R^2$ rather
than the distances $d_i$.

\medskip\noindent\textbf{Pratt fit}. To improve the performance of
the K{\aa}sa method one can minimize another function, $\cF =
\frac{1}{4R^2} \cF_{\rm K}$, which provides a better approximation
to $\sum d_i^2$. This new function, expressed in terms of $A,B,C,D$
reads
\beq \label{Pratt1}
  \cF_{\rm P} = \sum
    \frac{[Az_i+Bx_i+Cy_i+D]^2}{B^2+C^2-4AD},
\eeq
due to \eqref{conversion}. Equivalently, one can minimize
\beq \label{AAAA}
    \cF(A,B,C,D) = \sum
    [Az_i+Bx_i+Cy_i+D]^2
\eeq
subject to the constraint \eqref{constr}. This method was proposed
by Pratt \cite{Pr87}.

\medskip\noindent\textbf{Taubin fit}. A slightly different method
was proposed by Taubin \cite{Ta91} who minimizes the function
\beq
    \cF_{\rm T}=
    \frac{\sum \bigl[(x_i-a)^2 + (y_i-b)^2 - R^2\bigr]^2}
    {4n^{-1}\sum \bigl[(x_i-a)^2+(y_i-b)^2\bigr]}.
\eeq
Expressing it in terms of $A,B,C,D$ gives
\beq \label{Taubin1}
  \cF_{\rm T} = \sum
    \frac{[Az_i+Bx_i+Cy_i+D]^2}
    {n^{-1}\sum [4A^2z_i+4ABx_i+4ACy_i+B^2+C^2]}.
\eeq
Equivalently, one can minimize \eqref{AAAA} subject to a new
constraint
\beq \label{Tcon}
   4A^2\bar{z}+4AB\bar{x}+4AC\bar{y}+B^2+C^2=1.
\eeq
Here we use standard `sample means' notation: $\bar{x} = \frac 1n
\sum x_i$, etc.

\medskip\noindent\textbf{General remarks}.
Note that the minimization of \eqref{AAAA} must use some constraint,
to avoid a trivial solution $A=B=C=D=0$. Pratt and Taubin fits
utilize constraints \eqref{constr} and \eqref{Tcon}, respectively.
K{\aa}sa fit also minimizes \eqref{AAAA}, but subject to constraint
$A=1$.

While the Pratt and Taubin estimates of the parameters $A,B,C,D$
have finite moments, the corresponding estimates of $a,b,R$ have
infinite moments, just like the MLE \eqref{inf}. On the other hand,
K{\aa}sa's estimates of $a,b,R$ have finite moments whenever $n\geq
4$; see \cite{ZC06a}.

All the above circle fits have an important property -- they are
independent of the choice of the coordinate system, i.e.\ their
results are invariant under translations and rotations; see a proof
in \cite{web}.

Practical experience shows that the Pratt and Taubin fits are more
stable and accurate than the K{\aa}sa fit, and they perform nearly
equally well, see \cite{CL1}. Taubin \cite{Ta91} intended to compare
his fit to Pratt's theoretically, but no such analysis was ever
published. We make such a comparison below.

There are many other approaches to the circle fitting problem in the modern literature \cite{AW04,CS08,YW04,RTKD03,SSS03,SWFL00, UJ03,ZC06b,ZXA06}, but most of them are either quite slow or can be reduced to one of the algebraic fits \cite[Chapter~8]{web}.

\medskip\noindent\textbf{Matrix representation}.
We can represent the above three algebraic fits in matrix form. Let
$\bA=(A,B,C,D)$ denote the parameter vector,
\beq
     \bZ \stackrel{\rm def}= \left [
                  \begin{array}{cccc}
                     z_1 & x_1 & y_1 & 1 \\
                     \vdots & \vdots & \vdots & \vdots \\
                     z_n & x_n & y_n & 1
                     \end{array}
          \right]
\eeq
the `data matrix' (recall that $z_i = x_i^2+y_i^2)$ and
\beq \label{bM}
     \bM \stackrel{\rm def}= \frac 1n\, \bZ^T \bZ=
    \left [\begin{array}{cccc}
     \overline{zz} & \overline{zx} & \overline{zy} & \bar{z} \\
     \overline{zx} & \overline{xx} & \overline{xy} & \bar{x} \\
     \overline{zy} & \overline{xy} & \overline{yy} & \bar{y} \\
     \bar{z} & \bar{x} & \bar{y} & 1 \\
     \end{array}\right ]
\eeq
the `matrix of moments'. All the algebraic circle fits minimize the
same objective function ${\cF} (\bA) = \bA^T \bM \bA$, cf.\
\eqref{AAAA}, subject to a constraint $\bA^T \bN \bA = 1$, where the
matrix $\bN$ corresponds to the fit. The Pratt fit uses
 \beq \label{bNPratt}
   \bN=\bP\stackrel{\rm def}=\left [\begin{array}{rrrr}
     0 & 0 & 0 & -2 \\
     0 & 1 & 0 & 0 \\
     0 & 0 & 1 & 0 \\
     -2 & 0 & 0 & 0
     \end{array}\right ] ,
\eeq
the Taubin fit uses
\beq \label{bNTaubin}
   \bN=\bT\stackrel{\rm def}=\left [\begin{array}{cccc}
     4\bar{z} & 2\bar{x} & 2\bar{y} & 0 \\
     2\bar{x} & 1 & 0 & 0 \\
     2\bar{y} & 0 & 1 & 0 \\
     0 & 0 & 0 & 0
     \end{array}\right ],
\eeq
and the K{\aa}sa uses $\bN=\bK\stackrel{\rm def}=\be_1\be_1^T$, where
$\be_1=(1,0,0,0)^T$.

To solve the above constrained minimization problem one uses a
Lagrange multiplier $\eta$ and reduces it to unconstrained
minimization of the function
\beq
   \cG(\bA,\eta) = \bA^T \bM \bA
   - \eta (\bA^T \bN \bA - 1).
\eeq
Differentiating with respect to $\bA$ gives
\beq \label{MNeta}
  \bM \bA   = \eta \bN \bA,
\eeq
thus $\bA$ must be a generalized eigenvector for the matrix pair
$(\bM, \bN)$. This fact is sufficient for the subsequent analysis,
because it determines $\bA$ up to a scalar multiple, and multiplying
$\bA$ by a scalar does not change the circle it represents, so we
can set $\|\bA\| =1$.

\medskip\emph{Remark}. The generalized eigenvalue problem \eqref{MNeta} may have several solutions. To choose the right one we note that for each solution $(\eta,\bA)$
$$
  \bA^T \bM \bA   = \eta \bA^T \bN \bA = \eta,
$$
thus for the purpose of minimizing $\bA^T \bM \bA$ we should choose the solution of \eqref{MNeta} with the smallest positive $\eta$.

\section{Error Analysis: a general scheme} \label{SecEAGS}

We employ an error analysis scheme based on a `small noise'
assumption. That is, we assume that the errors $\delta_i$ and
$\eps_i$ (Section~\ref{SecSM}) are small and treat their standard
deviation $\sigma$ as a small parameter. The sample size $n$ is
fixed, though it is not very small.

This approach goes back to Kadane \cite{Ka70} and was employed by
Anderson \cite{An76} and other statisticians \cite{AS82}. More
recently it has been used by Kanatani \cite{Ka04a,Ka08a} in image
processing applications, who argued that the `small noise' model,
where $\sigma\to 0$ while the sample size $n$ is kept fixed, is more
appropriate than the traditional statistical `large sample'
approach, where $n\to \infty$ while $\sigma>0$ is kept fixed. We use
a combination of these two models: our main assumption is $\sigma\to
0$, but $n$ is regarded as a slowly increasing parameter; more precisely we assume $n\ll \sigma^{-2}$.

Suppose one is fitting curves defined by an implicit equation
\beq \label{Pcurve}
   P(x,y; \bTheta) = 0,
\eeq
where $\bTheta = (\theta_1, \ldots, \theta_k)^T$ denotes a vector of
unknown parameters to be estimated. Let $\tilde{\bTheta} =
(\tilde{\theta}_1, \ldots, \tilde{\theta}_k)^T$ be the 'true'
parameter vector corresponding to the `true' curve $P(x,y;
\tilde{\bTheta}) = 0$. As in Section~\ref{SecSM} let $(\tx_i,
\ty_i)$, $i = 1, \ldots, n$, denote true points, which lie on the
true curve, and $(x_i,y_i)$ observed points satisfying
\eqref{errors}. Let $\hat{\bTheta}(x_1,y_1, \ldots,x_n,y_n)$ be an
estimator. We assume that $\hat{\bTheta}$ is a regular (at least
four times differentiable) function of observations $(x_i, y_i)$.
The existence of the derivatives of $\hat{\bTheta}$ is only required
at the true points $(x_i,y_i)=(\tx_i,\ty_i)$, and it follows from
the implicit function theorem under general assumptions provided
$P(x,y; \bTheta)$ in \eqref{Pcurve} is differentiable (in most
cases $P$ is a polynomial in all its variables); we omit the proof.

For brevity we denote by $\bX = (x_1, y_1, \ldots, x_n, y_n)^T$ the
vector of all observations, so that $\bX = \tbX + \bE$, where $\tbX
= (\tx_1, \ty_1, \ldots, \tx_n, \ty_n)^T$ is the vector of the true
coordinates and $\bE = (\delta_1, \eps_1, \ldots, \delta_n,
\eps_n)^T$ is the `noise vector'; the components of $\bE$ are
i.i.d.\ normal random variables with mean zero and variance
$\sigma^2$.

We use Taylor expansion to the second order terms. To keep our
notation simple, we work with each scalar parameter $\theta_m$ of
the vector $\bTheta$ separately:
\beq \label{Texp2}
   \hat{\theta}_m (\bX) = \hat{\theta}_m (\tbX)
   + \bG_m^T \bE + \tfrac 12 \,
   \bE^T  \bH_m \bE
   + \cO_P(\sigma^3).
\eeq
Here $\bG_m = \nabla \hat{\theta}_m$ and $\bH_m = \nabla^2
\hat{\theta}_m$ denote the gradient (the vector of the first order
partial derivatives) and the Hessian matrix of the second order
partial derivatives of $\hat{\theta}_m$, respectively, taken at the
true vector $\tilde{\bX}$. The remainder term $\cO_P(\sigma^3)$ in
\eqref{Texp2} is a random variable $\cR$ such that $\sigma^{-3} \cR$
is bounded in probability.

Expansion \eqref{Texp2} shows that $\hat{\bTheta}(\bX) \to
\hat{\bTheta}(\tbX)$ in probability, as $\sigma \to 0$. It is
convenient to assume that
\beq \label{GeoCon}
   \hat{\bTheta} (\tbX) = \tbTheta.
\eeq
Precisely \eqref{GeoCon} means that whenever $\sigma=0$, i.e.\ when
the true points are observed without noise, then the estimator
returns the true parameter vector, i.e.\ finds the true curve.
Geometrically, it means that if there is a model curve that
interpolates the data points, then the algorithm finds it.

With some degree of informality, one can assert that whenever
\eqref{GeoCon} holds, the estimate $\hat{\bTheta}$ is
\emph{consistent} in the limit $\sigma \to 0$. This is regarded as a
minimal requirement for any sensible fitting algorithm. For example,
if the observed points lie on one circle, then every circle fitting
algorithm finds that circle uniquely. Kanatani \cite{Ka05b} remarks
that algorithms which fail to follow this property ``are not worth
considering''.

Under the assumption \eqref{GeoCon} we rewrite \eqref{Texp2} as
\beq \label{Texp2d}
   \Delta \hat{\theta}_m (\bX) =
   \bG_m^T \bE + \tfrac 12 \,
   \bE^T \bH_m \bE
   + \cO_P(\sigma^3),
\eeq
where $\Delta \hat{\theta}_m (\bX) = \hat{\theta}_m (\bX) -
\ttheta_m$ is the statistical error of the parameter estimate.

The accuracy of an estimator $\hat{\theta}$ in statistics is
characterized by its Mean Squared Error (MSE)
\beq
   \IE\bigl[(\hat{\theta}
  -\tilde{\theta})^2\bigr] =
  {\sf Var}(\hat{\theta}) +
  \bigr[{\rm bias}(\hat{\theta})\bigr]^2,
\eeq
where ${\rm bias}(\hat{\theta}) = \IE(\hat{\theta}) -
\tilde{\theta}$. But it often happens that exact (or even
approximate) values of $\IE(\hat{\theta})$ and ${\sf
Var}(\hat{\theta})$ are unavailable because the probability
distribution of $\hat{\theta}$ is overly complicated, which is
common in curve fitting problems, even if one fits straight lines to
data points; see \cite{An76, AS82}. There are also cases where the
estimates have theoretically infinite moments because of somewhat
heavy tails, which on the other hand barely affect their practical
performance. Thus their accuracy should not be characterized by the
theoretical moments which happen to be affected by heavy tails; see
also \cite{An76}. In all such cases one usually constructs a good
approximate probability distribution for $\hat{\theta}$ and judges
the quality of $\hat{\theta}$ by the moments of that distribution.

It is standard \cite{AF88, An76, AS82, Fubook, WF82} to construct a
normal approximation to $\hat{\theta}$ and treat its variance as an
`approximative' MSE of $\hat{\theta}$. The normal approximation
is usually based on the leading term in the Taylor expansion, like
$\bG_m^T \bE$ in \eqref{Texp2d}. For circle fitting algorithms, the resulting variance (see below) will be the same for all known methods, so we will go one step further and use the
second order term. This gives us a better approximative
distribution and allows us to compare circle fitting methods. In our formulas,
$\IE(\hat{\theta}_m)$ and ${\sf Var}(\hat{\theta}_m)$ denote the
mean and variance of the resulting approximative
distribution.

The first term in \eqref{Texp2d} is a linear
combination of i.i.d.\ normal random variables that have zero mean,
hence it is itself a normal random variable with zero mean. The
second term is a quadratic form of i.i.d.\ normal variables. Since
$\bH_m$ is a symmetric matrix, we have $\bH_m = \bQ_m^T \bD_m
\bQ_m$, where $\bQ_m$ is an orthogonal matrix and $\bD_m = \,{\rm
diag} \{d_1, \ldots, d_{2n}\}$ is a diagonal matrix. The vector
$\bE_m = \bQ_m \bE$ has the same distribution as $\bE$ does, i.e.\
its components are i.i.d.\ normal random variables with mean zero
and variance $\sigma^2$. Thus
\beq \label{chisqgen}
   \bE^T \bH_m \bE = \bE_m^T \bD_m \bE_m
   = \sigma^2 \sum d_i Z_i^2,
\eeq
where the $Z_i$'s are i.i.d.\ standard normal random variables, and
the mean value of \eqref{chisqgen} is
\beq
   \IE\bigl(\bE^T \bH_m \bE\bigr)
   =\sigma^2\,{\rm tr}\,\bD_m
   =\sigma^2\,{\rm tr}\,\bH_m.
\eeq
Therefore, taking the mean value in \eqref{Texp2d} gives
\beq \label{Tbias}
   {\rm bias}(\hat{\theta}_m) =
   \IE(\Delta \hat{\theta}_m) =
   \tfrac 12 \,
   \sigma^2\, {\rm tr}\, \bH_m
   + \cO(\sigma^4).
\eeq
Note that the expectations of all third order terms
vanish, because the components of $\bE$ are independent and their
first and third moments are zero; thus the remainder term is of
order $\sigma^4$.

Squaring \eqref{Texp2d} and again using \eqref{chisqgen} give the
mean squared error (MSE)
\beq \label{TMSE}
   \IE\bigl([\Delta \hat{\theta}_m]^2\bigr) =
   \sigma^2\bG_m^T\bG_m + \tfrac 14 \,
   \sigma^4 \bigl([{\rm tr}\, \bH_m]^2
   +2\|\bH_m\|_F^2\bigr)  + \cR,
\eeq
where $\|\bH_m\|_F^2 = \,{\rm tr}\, \bH_m^2$ is the Frobenius norm (note that $\|\bH_m\|_F^2 = \|\bD_m\|_F^2 =\,{\rm tr}\, \bD_m^2$).
The remainder $\cR$ includes terms of order $\sigma^6$, as well as
some terms of order $\sigma^4$ that contain third order partial
derivatives, such as $\partial^3\hat{\theta}_m/\partial x_i^3$ and
$\partial^3\hat{\theta}_m/\partial x_i^2\partial x_j$.
A similar expression can be derived for $\IE\bigl(\Delta \hat{\theta}_m\Delta \hat{\theta}_{m'}\bigr)$ for $m\neq m'$, we omit it and only give the final formula below.
\medskip

\noindent\textbf{Classification of higher order terms.} In the MSE
expansion \eqref{TMSE}, the leading term $\sigma^2\bG_m^T\bG_m$ is
the most significant. The terms of order $\sigma^4$ are often
given by long complicated formulas. Even the expression for the bias
\eqref{Tbias} may contain several terms of order $\sigma^2$, as we
will see below. Fortunately, it is possible to sort them out keeping
only the most significant ones, see next.

Kanatani \cite{Ka08a} recently derived formulas for the bias of
certain ellipse fitting algorithms. First he found all the terms of
order $\sigma^2$, but in the end he noticed that some terms were of
order $\sigma^2$ (independent of $n$), while the others of order
$\sigma^2/n$. The magnitude of the former was clearly larger than
that of the latter, and when Kanatani made his conclusions he
ignored the terms of order $\sigma^2/n$. Here we formalize
Kanatani's classification of higher order terms
as follows:\\
-- In the expression for the bias \eqref{Tbias} we keep terms of
order $\sigma^2$ (independent of $n$) and ignore terms of order
$\sigma^2/n$.\\ -- In the expression for the mean squared error
\eqref{TMSE} we keep terms of order $\sigma^4$ (independent of $n$)
and ignore terms of order $\sigma^4/n$.

These rules agree with our assumption that not
only $\sigma \to 0$, but also $n \to \infty$, although $n$ increases
rather slowly ($n \ll 1/\sigma^2$). Such models were studied by
Amemiya, Fuller and Wolter \cite{AF88, WF82} who made a more rigid assumption that $n
\sim \sigma^{-a}$ for some $0<a<2$.

Now it turns out (we omit detailed proofs; see \cite{web}) that the main term
$\sigma^2 \bG_m^T \bG_m$ in our expression for the MSE \eqref{TMSE}
is of order $\sigma^2/n$; so it will never be ignored. Of the fourth
order terms,  $\tfrac 12 \,
\sigma^4\|\bH_m\|_F^2$ is of order $\sigma^4/n$, hence it will be
discarded, and the same applies to all the terms
involving third order partial derivatives mentioned above.

The bias $\sigma^2\, {\rm tr}\, \bH_m$ in \eqref{Tbias} is,
generally, of order $\sigma^2$ (independent of $n$), thus its
contribution to the mean squared error \eqref{TMSE} is significant.
However the full expression for the bias may contain terms of order
$\sigma^2$ and of order $\sigma^2/n$, of which the latter will be
ignored; see below.

Now the terms in \eqref{TMSE} have the following orders of
magnitude:
\beq \label{TMSEO}
   \IE\bigl([\Delta \hat{\theta}_m]^2\bigr) =
   \cO(\sigma^2/n) + \cO(\sigma^4)
   +\cO(\sigma^4/n)  + \cO(\sigma^6),
\eeq
where each big-O simply indicates the order of the corresponding
term in \eqref{TMSE}. It is interesting to roughly compare their
values numerically. In typical computer vision applications,
$\sigma$ does not exceed $0.05$; see \cite{Be89}. The number of data
points normally varies between 10-20 (on the low end) and a few
hundred (on the high end). For simplicity, we can set $n \sim
1/\sigma$ for smaller samples and $n \sim 1/\sigma^2$ for larger
samples. Then Table~\ref{TabLHend} presents the corresponding
typical magnitudes of each of the four terms in \eqref{TMSE}.

\begin{table}
\begin{center}
\renewcommand{\arraystretch}{1.5}
\begin{tabular}{|c|cccc|}
\hline  & $\phantom{m}\sigma^2/n\phantom{m}$ &
        $\phantom{m}\sigma^4\phantom{m}$ & $\phantom{m}\sigma^4/n\phantom{m}$ &
        $\phantom{m}\sigma^6\phantom{m}$ \\
\hline  small samples ($n\sim 1/\sigma$) & $\sigma^3$ & $\sigma^4$ & $\sigma^5$ & $\sigma^6$ \\
        large samples ($n\sim 1/\sigma^2$) & $\sigma^4$ & $\sigma^4$ &
        $\sigma^6$ & $\sigma^6$ \\ \hline
\end{tabular}
\vspace{0.3cm} \caption{The order of magnitude of the four terms in
\eqref{TMSE}.} \label{TabLHend}
\end{center}
\end{table}

We see that for larger samples the fourth order term coming from the
bias may be just as big as the leading second-order term, hence it
would be unwise to ignore it. Earlier studies, see e.g.\ \cite{Be89,
CL2, Kabook}, usually focused on the leading, i.e.\ second-order,
terms only, disregarding all the fourth-order terms, and this is
where our analysis is different. We make one step further -- we keep
all the terms of order $\cO(\sigma^2/n)$ and $\cO(\sigma^4)$. The
less significant terms of order $\cO(\sigma^4/n)$ and
$\cO(\sigma^6)$ would be discarded.

Now combining all our results gives a matrix formula for the (total) mean squared error (MSE)
\beq \label{TMSET}
   \IE\Bigl[ (\Delta \hat{\bTheta})
   (\Delta \hat{\bTheta})^T\Bigr] =
   \sigma^2\bG\bG^T +
   \sigma^4 \bB \bB^T
    + \cdots,
\eeq
where $\bG$ is the $k\times 2n$ matrix of first order partial
derivatives of $\hat{\bTheta} (\bX)$, its rows are $\bG_m^T$, $1\leq
m\leq k$, and $\bB = \tfrac 12 [ \,{\rm tr}\, \bH_1, \ldots \,{\rm
tr}\, \bH_k]^T$ is the $k$-vector that represents the leading term
of the bias of $\hat{\bTheta}$, cf.\ \eqref{Tbias}. The trailing
dots in \eqref{TMSET} stand for all insignificant terms (those of
order $\sigma^4/n$ and $\sigma^6$).

We call the first (main) term $\sigma^2\bG\bG^T$ in \eqref{TMSET}
the \emph{variance} term, as it characterizes the variance (more
precisely, the covariance matrix) of the estimator $\hat{\bTheta}$,
to the leading order. For brevity we denote $\bV = \bG\bG^T$. The
second term $\sigma^4 \bB \bB^T$ is the `tensor square' of the bias
$\sigma^2 \bB$ of the estimator, again to the leading order. When we
deal with particular estimators in the next sections, we will see
that the actual expression for the bias is a sum of terms of two
types: some of them are of order $\cO(\sigma^2)$ and some others are
of order $\cO(\sigma^2/n)$, i.e.
\beq
   \IE(\Delta \hat{\bTheta}) = \sigma^2\bB
   +\cO(\sigma^4) = \sigma^2 \bB_1 + \sigma^2 \bB_2
   +\cO(\sigma^4),
\eeq
where $\bB_1 = \cO(1)$ and $\bB_2 = \cO(1/n)$. We call $\sigma^2
\bB_1$ the \emph{essential bias} of the estimator $\hat{\bTheta}$.
This is its bias to the leading order, $\sigma^2$. The other terms,
i.e.\ $\sigma^2 \bB_2$, and $\cO(\sigma^4)$, constitute
non-essential bias; they can be discarded. Now \eqref{TMSET} can be
written as
\beq \label{TMSET0}
   \IE\Bigl[ (\Delta \hat{\bTheta})
   (\Delta \hat{\bTheta})^T\Bigr] =
   \sigma^2\bV +
   \sigma^4 \bB_1 \bB_1^T
    + \cdots,
\eeq
where we only keep significant terms of order $\sigma^2/n$ and
$\sigma^4$ and drop the rest.

\medskip\noindent\textbf{KCR lower bound}.
The matrix $\bV$ representing the leading terms of the variance has
a natural lower bound (an analogue of the Cramer-Rao bound): for
every curve family \eqref{Pcurve} there is a symmetric positive
semi-definite matrix $\bV_{\min}$ such that for every estimator
satisfying \eqref{GeoCon}
\beq \label{KCR0}
   \bV \geq \bV_{\min} =
    \biggl(\sum
   \frac{P_{\bTheta i}\,P_{\bTheta i}^T}
   {\|P_{\bx i}\|^2}\biggr)^{-1},
\eeq
in the sense that $\bV - \bV_{\min}$ is a positive semi-definite
matrix. Here
\beq
   P_{\bTheta i}=\Bigl(\partial P(\tilde{\bx}_i;\tilde{\bTheta})/\partial \theta_1,
   \ldots,
   \partial P(\tilde{\bx}_i;\tilde{\bTheta})/\partial \theta_k\Bigr)^T
\eeq
stands for the gradient of $P$ with respect to the model parameters
$\theta_1, \ldots, \theta_k$ and
\beq
   P_{\bx i}=\Bigl(\partial P(\tilde{\bx}_i;\tilde{\bTheta})/\partial x,
   \partial P(\tilde{\bx}_i;\tilde{\bTheta})/\partial y\Bigr)^T
\eeq
for the gradient with respect to the planar variables $x$ and $y$;
both gradients are taken at the true point $\tbx_i = (\tx_i,
\ty_i)$. For example in the case of fitting circles defined by $P =
(x-a)^2+(y-b)^2-R^2$, we have
\beq
     P_{\bTheta i} =
     -2\bigl((\tilde{x}_i-\ta),(\tilde{y}_i-\tb),\tR\bigr)^T,
       \qquad
     P_{\bx i} =
     2\bigl((\tilde{x}_i-\ta),(\tilde{y}_i-\tb)\bigr)^T.
\eeq
Therefore,
\beq \label{bVmincir}
   \bV_{\min} = (\bW^T\bW)^{-1},
\eeq
where
\beq \label{bW}
     \bW \stackrel{\rm def}= \left [\begin{array}{ccc}
     \tu_1 & \tv_1 & 1 \\
     \vdots & \vdots & \vdots \\
     \tu_n & \tv_n & 1 \\
     \end{array}\right ]
\eeq
and $\tu_i, \tv_i$ are given by \eqref{tuv}.

The general inequality \eqref{KCR0} was proved by Kanatani
\cite{Kabook, Ka98} for unbiased estimators $\hat{\bTheta}$ and then
extended by Chernov and Lesort \cite{CL2} to all estimators
satisfying \eqref{GeoCon}. The geometric fit (which minimizes
orthogonal distances) always satisfies \eqref{GeoCon} and attains
the lower bound $\bV_{\min}$; this was proved by Fuller
(Theorem~3.2.1 in \cite{Fubook}) and independently by Chernov and
Lesort \cite{CL2}, who named the inequality \eqref{KCR0}
\emph{Kanatani-Cramer-Rao} (KCR) lower bound. See also survey \cite{Meer04} for the more general case of heteroscedastic noise.

\medskip\noindent\textbf{Assessing the quality of estimators}. Our
analysis dictates the following strategy of assessing the quality of
an estimator $\hat{\bTheta}$: first of all, its accuracy is
characterized by the matrix $\bV$, which must be compared to the KCR
lower bound $\bV_{\min}$. We will see that for all the circle
fitting algorithms the matrix $\bV$ actually achieves its lower
bound $\bV_{\min}$, i.e.\ we have $\bV= \bV_{\min}$, hence these
algorithms are optimal to the leading order.

Next, once the factor $\bV$ is already at its natural minimum, the
accuracy of an estimator should be characterized by the vector
$\bB_1$ representing the essential bias -- better estimates should
have smaller essential biases. It appears that there is no natural
minimum for $\|\bB_1\|$, in fact there exist estimators which have a
minimum variance $\bV = \bV_{\min}$ and a zero essential bias, i.e.\
$\bB_1 = \mathbf{0}$. We will construct such an estimator in
Section~\ref{SecEAACF}.


\section{Error analysis of geometric circle fit} \label{SecEAGCF}

Here we apply the general method of the previous section to the
geometric circle fit, i.e.\ to the estimator $\hat{\bTheta} =
(\hat{a}, \hat{b}, \hat{R})$ of the circle parameters minimizing the
sum $\sum d_i^2$ of orthogonal (geometric) distances from the data
points to the fitted circle.

\medskip\noindent\textbf{Variance of the geometric circle fit.}
We start with the main part of our error analysis -- the variance
term represented by $\sigma^2\bV$ in \eqref{TMSET0}. The distances
$d_i = r_i - R$ can be expanded as
\begin{align}
 d_i &= \sqrt{\bigl[(\tx_i+\delta_i)-(\ta+ \Delta a )]^2+  \bigl[(\ty_i+\eps_i)-(\tb+ \Delta b )]^2} - \tR-\Delta R \nonumber \\
     &=  \sqrt{\tR^2+2 \tR\tu_i (\delta_i-\Delta a)
        +2\tR\tv_i(\eps_i-\Delta b)+ \cO_P(\sigma^2)} - \tR-\Delta R  \nonumber\\
     &= \tu_i(\delta_i-\Delta a)
        +\tv_i (\eps_i-\Delta b)- \Delta R + \cO_P(\sigma^2),
\end{align}
see \eqref{tuv}. Minimizing $\sum d_i^2$ to the first order is
equivalent to minimizing
\beq
  \sum (\tu_i \, \Delta a + \tv_i \, \Delta b
  + \Delta R - \tu_i \delta_i - \tv_i \eps_i)^2.
\eeq
This is a classical least squares problem that can also be written as
\beq \label{WTappr}
   \bW \, \Delta \bTheta \approx \tbU  \bdelta + \tbV \beps,
\eeq
where $\bW$ is given by \eqref{bW}, $\bTheta = (a,b,R)^T$, as well
as $\bdelta = (\delta_1, \ldots, \delta_n)^T$ and $\beps = (\eps_1,
\ldots, \eps_n)^T$, while $\tbU = \,{\rm diag}(\tu_1, \ldots,
\tu_n)$ and $\tbV = \,{\rm diag}(\tv_1, \ldots, \tv_n)$. The
solution of the least squares problem \eqref{WTappr} is
\beq \label{DeltaThetaC1}
   \Delta \hat{\bTheta} = (\bW^T \bW)^{-1}
   \bW^T(\tbU  \bdelta + \tbV \beps),
\eeq
of course this does not include the $\cO_P(\sigma^2)$ terms. Thus
the variance of our estimator, to the leading order, is
\beq
  \IE\bigl[(\Delta \hat{\bTheta})(\Delta \hat{\bTheta})^T\bigr]
  = (\bW^T \bW)^{-1}
   \bW^T\IE\bigl[(\tbU  \bdelta + \tbV \beps)
   (\bdelta^T \tbU + \beps^T \tbV)\bigr]\bW(\bW^T \bW)^{-1}.
\eeq
Now observe that $ \IE(\bdelta \beps^T) = \IE(\beps \bdelta^T) =
\mathbf{0}$, as well as $\IE(\bdelta \bdelta^T) = \IE(\beps \beps^T)
=\sigma^2 \bI$, and we have $\tbU^2 + \tbV^2 = \bI$. Thus to the
leading order
\begin{align} \label{CirGeoVar}
  \IE\bigl[(\Delta \hat{\bTheta})(\Delta \hat{\bTheta})^T\bigr]
  = \sigma^2 (\bW^T \bW)^{-1}\bW^T\bW(\bW^T \bW)^{-1}
  = \sigma^2 (\bW^T \bW)^{-1},
\end{align}
where the higher order (of $\sigma^4$) terms are not included.
Comparing this to \eqref{bVmincir} confirms that the geometric fit
attains the minimal possible covariance matrix $\bV$.

\medskip\noindent\textbf{Bias of the geometric circle fit.}
Now we do a second-order error analysis, which has not been
previously done in the literature. According to a general formula
\eqref{Texp2}, we put
\beq
\begin{split}
         a &= \ta + \Delta_1 a + \Delta_2 a + \cO_P(\sigma^3),\\
         b &= \tb + \Delta_1 b + \Delta_2 b + \cO_P(\sigma^3),\\
         R &= \tR + \Delta_1 R + \Delta_2 R + \cO_P(\sigma^3).
\end{split}
\eeq
Here $\Delta_1 a$, $\Delta_1 b$, $\Delta_1 R$ are linear
combinations of $\eps_i$'s and $\delta_i$'s, which were found above,
in \eqref{DeltaThetaC1}, and $\Delta_2 a$, $\Delta_2 b$, $\Delta_2
R$ are quadratic forms of $\eps_i$'s and $\delta_i$'s to be
determined next.

Expanding the distances $d_i$ to the second order terms gives
\begin{align}
  d_i &= \tu_i(\delta_i-\Delta_1 a)
  +\tv_i (\eps_i-\Delta_1 b)- \Delta_1 R\nonumber\\
  &\quad  -\tu_i\,\Delta_2 a-\tv_i\,\Delta_2 b -\Delta_2 R
  +\tfrac{\tv_i^2}{2\tR}(\delta_i-\Delta_1 a)^2
  +\tfrac{\tu_i^2}{2\tR}(\eps_i-\Delta_1 b)^2\nonumber\\
  &\quad-\tfrac{\tu_i\tv_i}{\tR}(\delta_i-\Delta_1 a)(\eps_i-\Delta_1 b).
\end{align}
Since we already found $\Delta_1 a$, $\Delta_1 b$, $\Delta_1 R$, the
only unknowns are $\Delta_2 a$, $\Delta_2 b$, $\Delta_2 R$.
Minimizing $\sum d_i^2$ is now equivalent to minimizing
\beq
  \sum (\tu_i \, \Delta_2 a + \tv_i \, \Delta_2 b
  + \Delta_2 R - f_i)^2,
\eeq
where
\begin{eqnarray}
  f_i &=&\tu_i(\delta_i-\Delta_1 a)+\tv_i (\eps_i-\Delta_1 b)- \Delta_1 R \nonumber \\
   &+&  \tfrac{\tv_i^2}{2R}(\delta_i-\Delta_1 a)^2+\tfrac{\tu_i^2}{2R}(\eps_i-\Delta_1 b)^2-\tfrac{\tu_i\tv_i}{R}(\delta_i-\Delta_1 a)(\eps_i-\Delta_1 b).
   \label{fif}
\end{eqnarray}
This is another least squares problem, and its solution is
\beq \label{DeltaThetaC2}
   \Delta_2 \hat{\bTheta} = (\bW^T \bW)^{-1}
   \bW^T\bF,
\eeq
where $\bF = (f_1, \ldots, f_n)^T$; of course this is a quadratic
approximation which does not include $\cO_P(\sigma^3)$ terms. In
fact, the contribution from the first three (linear) terms in
\eqref{fif} vanishes, quite predictably; thus only the last two
(quadratic) terms matter.

Taking the mean value gives, to the leading order,
\beq \label{CirEBias}
  \IE(\Delta \hat{\bTheta}) =\IE(\Delta_2 \hat{\bTheta}) =
  \frac{\sigma^2}{2R}\bigl[
  (\bW^T \bW)^{-1} \bW^T \mathbf{1}
  +(\bW^T \bW)^{-1} \bW^T \bS\bigr],
\eeq
where $\mathbf{1} = (1,1,\ldots,1)^T$ and $\bS = (s_1, \ldots,
s_n)^T$, here $s_i$ is a scalar
\beq
   s_i = [-\tv_i, \tu_i, 0] (\bW^T \bW)^{-1} [-\tv_i, \tu_i,0]^T.
\eeq

The second term in (\ref{CirEBias}) is of order $\cO(\sigma^2/n)$,
thus the essential bias is given by the first term only, and it can
be simplified. Since the last column of the matrix $\bW^T \bW$
coincides with the vector $\bW^T \mathbf{1}$, we have $(\bW^T
\bW)^{-1} \bW^T \mathbf{1} = [0,0,1]^T$, hence the essential bias of
the geometric circle fit is
\beq \label{CirEBias001}
     \IE(\Delta \hat{\bTheta}) \stackrel{\rm ess}{=}
     \frac{\sigma^2}{2\tR}\, \bigl[0, 0, 1\bigr]^T.
\eeq
Thus the estimates of the circle center, $\hat{a}$
and $\hat{b}$, have \emph{no essential bias}, while the estimate of
the radius has essential bias
\beq \label{CirEBiasR}
     \IE(\Delta \hat{R}) \stackrel{\rm ess}{=} \frac{\sigma^2}{2\tR},
\eeq
which is independent of the number and location of the true points.
These facts are consistent with the results obtained by Berman
\cite{Be89} under the assumptions that $\sigma>0$ is fixed and
$n\to \infty$.


\section{Error analysis of algebraic circle fits} \label{SecEAACF}

Here we analyze algebraic circle fits using their matrix
representation.

\medskip\noindent\textbf{Matrix perturbation method.}
For every random variable, matrix or vector, $\bL$, we write
\beq \label{bZexp2}
   \bL = \tilde{\bL} + \Delta_1\bL + \Delta_2\bL +
   \cO_P(\sigma^3),
\eeq
where $\tilde{\bL}$ is its `true', nonrandom, value (achieved when
$\sigma=0$), $\Delta_1\bL$ is a linear combination of $\delta_i$'s
and $\eps_i$'s, and $\Delta_2\bL$ is a quadratic form of
$\delta_i$'s and $\eps_i$'s; all the higher order terms (cubic etc.)
are represented by $\cO_P(\sigma^3)$. For brevity, we drop the
$\cO_P(\sigma^3)$ terms in our formulas. Therefore $\bA =
\tilde{\bA} + \Delta_1\bA + \Delta_2\bA$ and $\bM = \tilde{\bM} +
\Delta_1\bM + \Delta_2\bM$, and \eqref{bM} implies
\begin{align}
  \Delta_1\bM&=n^{-1}(\tilde{\bZ}^T\Delta_1\bZ+\Delta_1\bZ^T\tilde{\bZ}),
  \label{linear1} \\
  \Delta_2 \bM &=n^{-1}(\Delta_1\bZ^T \, \Delta_1 \bZ
+ \tbZ^T \, \Delta_2 \bZ + \Delta_2\bZ^T \, \tbZ).\label{quadratic1}
\end{align}
Since the true points lie on the true circle, $\tbZ \tbA =
\mathbf{0}$, as well as  $\tbM \tbA = \mathbf{0}$ (hence $\tbM$ is a
singular matrix). Therefore
\beq
   \tbA^T \, \Delta_1 \bM \, \tbA =
   n^{-1}\tbA^T \bigl(\tbZ^T \, \Delta_1 \bZ + \Delta_1 \bZ^T \, \tbZ\bigr) \tbA
   =0,
\eeq
hence $\bA^T \bM \bA = \cO_P(\sigma^2)$, and premultiplying
(\ref{MNeta}) by $\bA^T$ yields $\eta = \cO_P (\sigma^2)$. Next,
substituting the expansions of $\bM$, $\bA$, and $\bN$ into
(\ref{MNeta}) gives
\beq
  (\tbM + \Delta_1 \bM + \Delta_2 \bM)
  (\tbA + \Delta_1 \bA + \Delta_2 \bA)
  = \eta \tbN \tbA
\eeq
(recall that $\bN$ is data-dependent for the Taubin method, but only
its `true' value $\tbN$ matters, as $\eta = \cO_P (\sigma^2)$, hence the use of the observed values only adds higher order terms). Now
using $\tbM\tbA=\b0$ yields
\beq \label{TexpMNeta}
( \tbM\, \Delta_1 \bA+ \Delta_1 \bM\, \tbA)
 + (\tbM \, \Delta_2 \bA+ \Delta_1 \bM \, \Delta_1 \bA + \Delta_2 \bM \, \tbA)
  = \eta \tbN \tbA
\eeq
The left hand side of \eqref{TexpMNeta} consists of a linear part
$(\tbM\, \Delta_1\bA + \Delta_1\bM\, \tbA)$ and a quadratic part
$(\tbM \, \Delta_2 \bA+ \Delta_1 \bM \, \Delta_1 \bA + \Delta_2 \bM
\, \tbA)$. Separating them gives
\beq \label{TexpMNeta1}
 \tbM\, \Delta_1\bA + n^{-1}\tbZ^T \, \Delta_1 \bZ\, \tbA = \mathbf{0}
\eeq
(where we used \eqref{linear1} and $\tbZ\tbA={\mathbf 0}$) and
\beq \label{TexpMNeta2}
  \tbM \, \Delta_2 \bA+ \Delta_1 \bM \, \Delta_1 \bA
  + \Delta_2 \bM \, \tbA= \eta \tbN \tbA.
\eeq
Note that $\tbM$ is a singular matrix (because $\tbM \tbA =
\mathbf{0}$), but whenever there are at least three distinct true
points, they determine a unique true circle, thus the kernel of
$\tbM$ is one-dimensional, and it coincides with span($\tbA$). Also,
we set $\| \bA \|=1$, hence $\Delta_1 \bA$ is orthogonal to $\tbA$,
and we can write
\beq \label{Delta1bA}
   \Delta_1\bA  = -n^{-1} \tbM^- \tbZ^T \, \Delta_1 \bZ\, \tbA,
\eeq
where $\tbM^-$ denotes the Moore-Penrose pseudoinverse. Now one can
easily check that $\IE(\Delta_1 \bM \, \Delta_1 \bA) =
\cO(\sigma^2/n)$ and $\IE(\Delta_1\bA)=0$; these facts will be
useful in the upcoming analysis.

\medskip\noindent\textbf{Variance of algebraic circle fits.}
From (\ref{Delta1bA}) we conclude that
\begin{align} \label{Delta1bAAT}
  \IE\bigl[(\Delta_1\bA)(\Delta_1\bA)^T\bigr] &=
  n^{-2}\tbM^- \IE(\tbZ^T\, \Delta_1\bZ \,\tbA \tbA^T\,
  \Delta_1 \bZ^T\, \tbZ) \tbM^-\nonumber\\
   &=
  n^{-2}\tbM^- \IE\Bigl[\bigl(\sum_i\tbZ_i\, \Delta_1\bZ_i^T\bigr)\tbA \tbA^T
  \bigl(\sum_j\Delta_1 \bZ_j^T\, \tbZ_j^T\bigr)\Bigr] \tbM^-,
\end{align}
where
\beq
  \tbZ_i \stackrel{\rm def}=\left [\begin{array}{c}
     \tz_i\\
     \tx_i\\
     \ty_i\\
     1
     \end{array}\right ]
     \qquad\text{and}\qquad
  \Delta_1 \bZ_i =\left [\begin{array}{c}
     2\tx_i\delta_i+2\ty_i\eps_i\\
     \delta_i\\
     \eps_i\\
     0
     \end{array}\right ]
\eeq
denote the columns of the matrices $\tbZ^T$ and $\Delta_1 \bZ^T$,
respectively. Next,
\beq
    \IE\bigl[(\Delta_1\bZ_i)(\Delta_1\bZ_j)^T\bigr]=
      \left \{ \begin{array}{cc}
         0 & \qquad\text{whenever }\quad i\neq j\\
 \sigma^2 \tbT_i  & \qquad\text{whenever }\quad i = j
     \end{array}       \right.
\eeq
where
\beq
  \tbT_i \stackrel{\rm def}=  \left [\begin{array}{cccc}
     4\tz_i & 2\tx_i & 2\ty_i & 0 \\
     2\tx_i & 1 & 0 & 0 \\
     2\ty_i & 0 & 1 & 0 \\
     0 & 0 & 0 & 0
     \end{array}\right ].
\eeq
Note $n^{-1}\sum \tbT_i=\tbT$ and $\tbA^T \tbT_i \tbA = \tbA^T \bP
\tbA = \tB^2+\tC^2-4\tA\tD $ for each $i$; recall \eqref{bNPratt} and \eqref{bNTaubin}. Hence
\beq
   \sum \tbZ_i \tbA^T \tbT_i \tbA \tbZ_i^T
   =\sum (\tbA^T  \bP \tbA) \tbZ_i  \tbZ_i^T
   = n(\tbA^T  \bP \tbA) \tbM.
\eeq
Combining the above formulas gives
\begin{align}
  \IE\bigl[(\Delta_1\bA)(\Delta_1\bA)^T\bigr]
   &=  n^{-2}\tbM^- \Bigl[\sum_{i,j} \tbZ_i \tbA^T \IE\bigl(\Delta_1\bZ_i^T
  \Delta_1 \bZ_j^T\bigr) \tbA \tbZ_j^T\Bigr] \tbM^- \nonumber\\
   &=   n^{-2}\sigma^2 \tbM^- \Bigl[\sum \tbZ_i \tbA^T
  \tbT_i \tbA \tbZ_i^T\Bigr] \tbM^- \nonumber \\
    &= n^{-1}\sigma^2 \tbM^-(\tbA^T  \bP \tbA). \label{CirCovA}
\end{align}
Remarkably, the variance of algebraic fits does not depend on the
constraint matrix $\bN$, hence all algebraic fits have the same
variance (to the leading order). In the next section we will derive
the variance of algebraic fits in the natural circle parameters
$(a,b,R)$ and see that it coincides with the variance of the
geometric fit (\ref{CirGeoVar}).

\medskip\noindent\textbf{Bias of algebraic circle fits.}
Since $\IE(\Delta_1\bA)=0$, it will be enough to find
$\IE(\Delta_2\bA)$. Premultiplying (\ref{TexpMNeta2}) by $\tbA^T$
yields
\beq \label{eta}
  \eta = \frac{\tbA^T \bM \bA}{\tbA^T \bN \bA} =
   \frac{\tbA^T\, \Delta_2\bM \, \tbA + \tbA^T \,
   \Delta_1 \bM \, \Delta_1\bA}{\tbA^T \tbN \tbA}
   +\cO_P(\sigma^2/n).
\eeq
Recall that $\IE(\Delta_1 \bM \, \Delta_1 \bA) = \cO(\sigma^2/n)$,
thus this term will not affect the essential bias and we drop it.
Taking the mean value and using (\ref{Delta1bA}) gives
\beq \label{Eeta}
  \IE(\eta) =
   \frac{\tbA^T \IE(\Delta_2 \bM) \tbA}
   {\tbA^T \tbN \tbA} + \cO(\sigma^2/n),
\eeq
We substitute \eqref{quadratic1} into \eqref{eta}, use $\tbZ \tbA =
{\mathbf 0}$, then observe that
\beq
  \IE(\Delta_1 \bZ^T\, \Delta_1 \bZ \tbA) =
  \sigma^2\sum\tbT_i\tbA = 2\tA\sigma^2\sum\tbZ_i + n\sigma^2\bP\tbA
  \label{CAbias4}
\eeq
(here $\tA$ is the first component of the vector $\tbA$). Then, note
that $\Delta_2 \bZ_i = (\delta_i^2+\eps_i^2, 0, 0, 0)^T$, and so
\beq \label{CAbias3}
   \IE(\tbZ^T\, \Delta_2 \bZ) \tbA
   =2\tA\sigma^2\sum\tbZ_i.
\eeq
Therefore the essential bias is given by
\beq \label{CirAlgBias}
   \IE(\Delta_2 \bA) \stackrel{\rm ess}{=} -\sigma^2 \tbM^-
   \Bigl[4\tA n^{-1}\sum\tbZ_i
   + \bP\tbA -
   \frac{\tbA^T \bP \tbA}{\tbA^T \tbN \tbA}\,\tbN\tbA\Bigr].
\eeq
A more detailed analysis (which we omit) gives the following
expression containing all the $\cO(\sigma^2)$ and $\cO(\sigma^2/n)$
terms:
\begin{align} \label{CirAlgBiasFull}
   \IE(\Delta_2 \bA) &= -\sigma^2 \tbM^-\Bigl[4\tA n^{-1}\sum\tbZ_i
   + (1-\tfrac 4n)\bP\tbA -
   \frac{(1-\tfrac 3n)(\tbA^T \bP \tbA)}{\tbA^T \tbN \tbA}\,\tbN\tbA
   \nonumber\\
   &\quad -4\tA n^{-2}\sum (\tbZ_i^T \tbM^-\tbZ_i)\tbZ_i\Bigr]
    + \cO(\sigma^4) .
\end{align}
This expression demonstrates that the terms of order $\sigma^2/n$
(the \emph{non-essential bias}) only add a small correction, which
is negligible when $n$ is large.

The expressions \eqref{CirAlgBias} and \eqref{CirAlgBiasFull} can be
simplified. Note that the vector $n^{-1}\sum\tbZ_i$ coincides with
the last column of the matrix $\tbM$, hence
\beq
    -\sigma^2 \tbM^-
   \Bigl[4\tA n^{-1}\sum\tbZ_i\Bigr] =-4\sigma^2\tA
   \,[0,0,0,1]^T.
\eeq
In fact, this term will play the key role in the subsequent
analysis.

\section{Comparison of various circle fits} \label{SecCVCF}

\medskip\noindent\textbf{Bias of the Pratt and Taubin fits.}
We have seen that all the algebraic fits have the same main
characteristic -- the variance (\ref{CirCovA}), to the leading
order. We will see below that their variance coincides with that of
the geometric circle fit. Thus the difference between all our circle
fits should be traced to the higher order terms, especially to their
essential biases.

First we compare the Pratt and Taubin fits. For the Pratt fit, the
constraint matrix is $\bN = \tbN = \bP$, hence its essential bias
\eqref{CirAlgBias} becomes
\beq \label{CirAlgBiasP}
   \IE(\Delta_2 \bA_{\rm Pratt})  \stackrel{\rm ess}{=}
    -4\sigma^2 \tA \,[0,0,0,1]^T.
\eeq
In other words, the Pratt constraint $\bN = \bP$ cancels the second
(middle) term in \eqref{CirAlgBias}; it leaves the first term
intact.

For the Taubin fit, the constraint matrix is $\bN = \bT$ and its
`true' value is $\tbN = \tbT = \tfrac 1n \sum \tbT_i$; also note
that $\tbT_i \tbA = 2\tA\tbZ_i + \bP \tbA$ for every $i$. Hence the
Taubin's bias is
\beq \label{CirAlgBiasT}
   \IE(\Delta_2 \bA_{\rm Taubin})  \stackrel{\rm ess}{=}
    -2\sigma^2 \tA \,[0,0,0,1]^T.
\eeq
Thus, the Taubin constraint $\bN = \bT$ cancels the second term in
\eqref{CirAlgBias} \emph{and} a half of the first term; it leaves
only a half of the first term in place.

As a result, the Taubin fit's essential bias is twice as small as
that of the Pratt fit. Given that their main terms (variances) are
equal, we see that the Taubin fit is statistically more accurate
than that of Pratt. We believe our analysis answers the question
posed by Taubin \cite{Ta91} who intended to compare his fit to
Pratt's.

\medskip\noindent\textbf{`Hyperaccurate' algebraic fit.}
Our error analysis leads to another stunning discovery -- an
algebraic fit that has \textit{no essential bias at all}. To our
knowledge, this is the first such algorithm for curve fitting
problems.

Let us set the constraint matrix to
\beq \label{bNHyper}
   \bN = \bH\stackrel{\rm def}=2\bT-\bP=\left [\begin{array}{cccc}
     8\bar{z} & 4\bar{x} & 4\bar{y} & 2 \\
     4\bar{x} & 1 & 0 & 0 \\
     4\bar{y} & 0 & 1 & 0 \\
     2 & 0 & 0 & 0
     \end{array}\right ].
\eeq
Then one can easily see that $\bH \tbA = 4\tA\,\tfrac{1}{n}\,\sum
\tbZ_i + \bP \tbA$, as well as $\tbA^T \bH \tbA = \tbA^T \bP \tbA$,
hence all the terms in (\ref{CirAlgBias}) cancel out! The resulting
essential bias vanishes:
\beq \label{CirAlgBiasH}
   \IE(\Delta_2 \bA_{\rm Hyper}) \stackrel{\rm ess}{=} 0.
\eeq
We call this fit \emph{hyperaccurate}, or `Hyper' for short. The
term \emph{hyperaccuracy} was introduced by Kanatani \cite{Ka06a,
Ka08a} who was first to employ Taylor expansion up to the terms of
order $\sigma^4$ for the purpose of comparing various algebraic fits
and designing better fits.

We note that the Hyper fit is invariant under translations and
rotations because its constraint matrix $\bH$ is a linear
combination of two others, $\bT$ and $\bP$, that satisfy the
invariance requirements; see a proof in \cite{web}.

As any other algebraic circle fit, the Hyper fit minimizes the function $\cF(\bA) = \bA^T \bM \bA$ subject to the constraint $\bA^T \bN \bA=1$ (with $\bN=\bH$), hence we need to solve the generalized eigenvalue problem $\bM \bA = \eta \bH \bA$ and choose the solution with the smallest positive eigenvalue $\eta$ (see the end of Section~\ref{SecACF}).

The matrix $\bH$ is not singular, three of its eigenvalues are positive and one is negative (these facts can be easily derived from the following simple observations: det$\,\bH=-4$, trace$\,\bH=8\bar{z}+2>1$, and $\lambda=1$ is one of its eigenvalues). Assume that $\bM$ is positive definite, then by Sylvester's law of inertia,
the matrix $\bH^{-1}\bM$ has the same signature as
$\bH$ does, i.e.\ the eigenvalues $\eta$ of $\bH^{-1} \bM$ are
all real, exactly three of them are positive and one is negative. The eigenpair $(\eta, \bA)$ with the negative eigenvalue $\eta$ does not represent any circle \cite{web}, so it is useless. (We note that Pratt's fit has similar properties, as det$\,\bP=-4$.) The eigenpair with the smallest positive $\eta$ gives the best fit. Lastly, the matrix $\bM$ is singular if and only if the observed points lie on a circle (or a line), in this case the eigenvector $\bA$ corresponding to $\eta=0$ gives the interpolating circle (line).

The Hyper fit can be computed by a numerically stable procedure
involving singular value decomposition (SVD). First, we compute the
(short) SVD, $\bZ = \bU \bSigma \bV^T$, of the matrix $\bZ$. If its
smallest singular value, $\sigma_{4}$, is less than a predefined
tolerance $\varepsilon$ (we suggest $\varepsilon = 10^{-12}$), then
$\bA$ is the corresponding right singular vector, i.e.\ the fourth
column of the $\bV$ matrix. In the regular case ($\sigma_{4} \geq
\varepsilon$), one forms $\bY = \bV \bSigma \bV^T$ and finds the
eigenpairs of the symmetric matrix $\bY \bH^{-1} \bY$. Selecting the
eigenpair $(\eta, \bA_\ast)$ with the smallest positive eigenvalue
and computing $\bA = \bY^{-1} \bA_\ast$ completes the solution. The prior translation of the coordinate system to the centroid of the data set (which ensures that $\bar{x} = \bar{y} = 0$) makes the computation of $\bH^{-1}$ particularly simple. The
corresponding MATLAB code is available from our web page \cite{web}.

\medskip\noindent\textbf{Transition between parameter schemes.}
Our next goal is to express the covariance and the essential bias of
the algebraic circle fits in terms of the natural parameters
$\bTheta = (a,b,R)^T$. Taking partial derivatives in
\eqref{conversion} gives a $3\times 4$ `Jacobian' matrix
\beq
   \bJ \stackrel{\rm def}= \left [\begin{array}{cccc}
     \frac{B^2}{2A^2} & -\frac{1}{2A} & 0 & 0 \\
     \frac{C^2}{2A^2} & 0 & -\frac{1}{2A} & 0 \\
     -\frac{R}{A}-\frac{D}{2A^2R} & \frac{B}{4A^2R} & \frac{C}{4A^2R} & -\frac{1}{2AR}
     \end{array}\right ].
\eeq
Thus we have
\beq \label{Delta1bThetaA}
   \Delta_1 \bTheta = \tbJ \, \Delta_1 \bA
   \qquad\text{and}\qquad
   \Delta_2 \bTheta = \tbJ \, \Delta_2 \bA
   + \cO_P(\sigma^2/n),
\eeq
where $\tbJ$ denotes the matrix $\bJ$ at the true parameters
$(\tA,\tB,\tC,\tD)$. The remainder term $\cO_P(\sigma^2/n)$ comes
from the second order partial derivatives, for example
\beq \label{abiasABCD}
  \Delta_2 a = (\nabla a)^T (\Delta_2 \bA) + \tfrac 12
   (\Delta_1 \bA)^T (\nabla^2 a) (\Delta_1 \bA),
\eeq
where $\nabla^2 a$ is the Hessian matrix of the second order partial
derivatives of $a$ with respect to $(A,B,C,D)$. The last term in
\eqref{abiasABCD} can be actually discarded, as it is of order
$\cO_P(\sigma^2/n)$ because $\Delta_1 \bA = \cO_P(\sigma/\sqrt{n})$.
We collect all such terms in the remainder term $\cO_P(\sigma^2/n)$
in \eqref{Delta1bThetaA}.

Next we need a useful fact. Suppose a point $(x_0,y_0)$ lies on the
true circle $(\ta, \tb, \tR)$, i.e.
\beq
   (x_0 - \ta)^2 + (y_0 - \tb)^2 = \tR^2.
\eeq
In accordance with our early notation we denote $z_0 = x_0^2 +
y_0^2$ and $\bZ_0 = (z_0, x_0, y_0, 1)^T$. We also put $u_0 =
(x_0-\ta)/\tR$ and $v_0 = (y_0-\tb)/\tR$, and consider the vector
$\bW_0 = (u_0, v_0, 1)^T$. The following formula will be useful:
\beq \label{FactMyst}
  2\tA \tR \tbJ \tbM^-\bZ_0 = -n(\bW^T \bW)^{-1} \bW_0,
\eeq
where the matrix $(\bW^T \bW)^{-1}$ appears in \eqref{CirEBias} and
the matrix $\tbM^-$ appears in \eqref{CirCovA}. The identity
\eqref{FactMyst} is easy to verify  directly for the unit circle
$\ta=\tb=0$ and $\tR=1$, and then one can check that it remains
valid under translations and similarities.

Equation \eqref{FactMyst} implies that for every true point $(\tx_i,
\ty_i)$
\beq
 4\tA^2\tR^2 \tbJ \tbM^-\tbZ_i\tbZ_i^T\tbM^-\tbJ^T =
 n^2(\bW^T \bW)^{-1} \bW_i\bW_i^T(\bW^T \bW)^{-1},
\eeq
where $\bW_i = (\tu_i, \tv_i, 1)^T$ denote the columns of the matrix
$\bW$, cf.\ \eqref{bW}. Summing up over $i$ gives
\beq \label{FactMyst1}
 4\tA^2\tR^2 \tbJ \tbM^-\tbJ^T =
 n(\bW^T \bW)^{-1}.
\eeq

\medskip\noindent\textbf{Variance and bias of algebraic
circle fits in the natural parameters.} Now we can compute the
variance (to the leading order) of the algebraic fits in the natural
geometric parameters. Notice that the third relation in
(\ref{conversion}) implies
$\tbA^T\bP\tbA=\tB^2+\tC^2-4\tA\tD=4\tA^2\tR^2$. Thus using
(\ref{FactMyst1}) gives
\begin{eqnarray} \label{CirAlgVara}
    \IE\bigl[(\Delta_1\bTheta)(\Delta_1\bTheta)^T\bigr]
    &=&\IE\bigl[\bJ(\Delta_1\bA)(\Delta_1\bA)^T\bJ^T\bigr] \nonumber\\
    &=&n^{-1} \sigma^2(\tbA^T  \bP \tbA)(\bJ\tbM^-\bJ^T) \nonumber\\
    &=&\sigma^2( 4\tA^2\tR^2n^{-1}\bJ\tbM^-\bJ^T) \nonumber \\
      &=& \sigma^2 (\bW^T \bW)^{-1}.
\end{eqnarray}
Thus the variance of all the algebraic circle fits (to the leading
order) coincides with that of the geometric circle fit, cf.\
(\ref{CirGeoVar}). Therefore the difference between all the circle
fits should be then characterized in terms of their biases, which we
do next.

The essential bias of the Pratt fit is, due to \eqref{CirAlgBiasP},
\index{Pratt circle fit}
\beq \label{CirEBiasP}
  \IE(\Delta_2 \hat{\bTheta}_{\rm Pratt}) \stackrel{\rm ess}{=}
   2\sigma^2\tR^{-1} \bigl[0, 0, 1\bigr]^T.
\eeq
Observe that the estimates of the circle center are essentially
unbiased, and the essential bias of the radius estimate is
$2\sigma^2/\tR$, which is independent of the number and location of
the true points. We know that the essential bias
\index{Taubin circle fit}
of the Taubin fit is twice as small, hence
\beq \label{CirEBiasT}
  \IE(\Delta_2 \hat{\bTheta}_{\rm Taubin}) \stackrel{\rm ess}{=}
    \sigma^2\tR^{-1} \bigl[0, 0, 1\bigr]^T.
\eeq
Comparing to \eqref{CirEBias001} shows that the geometric fit has an
essential bias that is twice as small as that of Taubin and four
times smaller than that of Pratt. Therefore, the geometric fit has
the smallest bias among all the popular circle fits, i.e.\ it is
statistically most accurate.

The formulas for the bias of the K{\aa}sa fit can be derived, too,
but in general they are complicated. However recall that all our
fits, including K{\aa}sa, are independent of the choice of the
coordinate system, hence we can choose it so that the true circle
has center at $(0,0)$ and radius $\tR=1$. For this circle $\tbA =
\frac{1}{\sqrt{2}}\, [1,0,0,-1]^T$, hence $\bP \tbA = 2\tbA$ and so
$\tbM^- \bP \tbA = {\mathbf 0}$, i.e.\ the middle term in
\eqref{CirAlgBias} is gone. Also note that $\tbA^T \bP \tbA = 2$,
hence the last term in parentheses in \eqref{CirAlgBias} is
$2\sqrt{2}\,[1,0,0,0]^T$.

\begin{figure}[htb]
    \centering
   \includegraphics[width=1.5in]{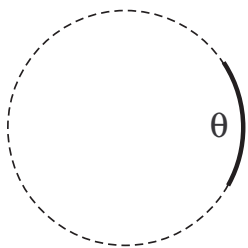}
    \caption{The arc containing the true points.}
    \label{FigArc}
\end{figure}

Next, assume for simplicity that the true points are equally spaced
on an arc of size $\theta$ (a typical arrangement in many studies).
Choosing the coordinate system so that the east pole $(1,0)$ is at
the center of that arc (see Figure~\ref{FigArc}) ensures $\bar{y} = \overline{xy}=0$. It is
not hard to see now that
\beq
   \tbM^-[1,0,0,0]^T = \tfrac 14 (\overline{xx}- \bar{x}^2)^{-1}
   [\overline{xx}, -2\bar{x}, 0,
   \overline{xx}]^T.
\eeq
Using the formula \eqref{Delta1bThetaA} we obtain (omitting details
as they are not so relevant) the essential bias of the K{\aa}sa fit
in the natural parameters $(a,b,R)$:
\beq
  \IE(\Delta_2 \hat{\bTheta}_{\rm Kasa}) \stackrel{\rm ess}{=}
   2\sigma^2 \bigl[0, 0, 1\bigr]^T\,-\,
  \frac{\sigma^2}{\overline{xx}- \bar{x}^2}
    \bigl[-\bar{x}, 0, \overline{xx}\bigr]^T.
\eeq
The first term here is the same as in \eqref{CirEBiasP} (recall that
$\tR=1$), but it is the second term above that causes serious
trouble: it grows to infinity because $\overline{xx}- \bar{x}^2 \to
0$ as $\theta \to 0$. This explains why the K{\aa}sa fit develops a
heavy bias toward smaller circles when data points are sampled from
a small arc.

\section{Experimental tests and conclusions}

To illustrate our analysis of various circle fits we have run a few
computer experiments where we set $n$ true points equally spaced
along a semicircle of radius $R=1$. Then we generated random samples
by adding a Gaussian noise at level $\sigma=0.05$ to each true
point, and after that applied various circle fits to estimate the
parameters $(a,b,R)$.

\begin{table}
\begin{center}
\renewcommand{\arraystretch}{1.5}
\begin{tabular}{|l|cccc|}
\hline  & \multicolumn{4}{c|}{total MSE = variance +
        (ess.\ bias$)^2$ + rest of MSE} \\
\hline  Pratt & \phantom{$m$}1.5164 & \phantom{$mm$}1.2647 & \phantom{$mm$}0.2500 & \phantom{$-$}0.0017 \\
        Taubin & \phantom{$m$}1.3451 & \phantom{$mm$}1.2647 & \phantom{$mm$}0.0625 & \phantom{$-$}0.0117 \\
        Geom. & \phantom{$m$}1.2952 & \phantom{$mm$}1.2647 & \phantom{$mm$}0.0156 & \phantom{$-$}0.0149 \\
        Hyper. & \phantom{$m$}1.2892 & \phantom{$mm$}1.2647 & \phantom{$mm$} 0.0000 & \phantom{$-$}0.0244 \\
\hline
\end{tabular}
\vspace{0.3cm} \caption{Mean square error (and its components) for
four circle fits ($10^4 \times$values are shown). In this test
$n=100$ points are placed (equally spaced) along a semicircle of
radius $R=1$ and the noise level is $\sigma = 0.05$.}
\label{Taballcir1}
\end{center}
\end{table}

Table~\ref{Taballcir1} summarizes the results of the first test,
with $n=100$ points; it shows the mean square error (MSE) of the
radius estimate $\hat{R}$ for each circle fit (obtained by averaging
over $10^7$ randomly generated samples). The table also gives the
breakdown of the MSE into three components. The first two are the
variance (to the leading order) and the square of the essential
bias, both computed according to our theoretical formulas. These two
components do not account for the entire mean square error, due to
higher order terms which our analysis discarded. The remaining part
of the MSE is shown in the last column, which is relatively small.
(We note that only the total MSE can be observed in practice; all the other columns of this table are the results of our theoretical analysis.)

We see that all the circle fits have the same (leading) variance,
which accounts for the `bulk' of the MSE. Their essential bias is
different, it is highest for the Pratt fit and smallest (zero) for
the Hyper fit. Algorithms with smaller essential biases perform
overall better, i.e.\ have smaller mean square error. The Hyper fit
is the best in our experiment; it outperforms the (usually
unbeatable) geometric fit.

\begin{table}
\begin{center}
\renewcommand{\arraystretch}{1.5}
\begin{tabular}{|l|rrrr|}
\hline  & \multicolumn{4}{c|}{total MSE = variance +
        (ess.\ bias$)^2$ + rest of MSE} \\
\hline  Pratt & \phantom{$m$}25.5520 & \phantom{$mm$}1.3197 & \phantom{$mm$}25.0000 & \phantom{$-$}-0.76784 \\
        Taubin & \phantom{$m$}7.4385 & \phantom{$mm$}1.3197 & \phantom{$mm$}6.2500 & \phantom{$-$}-0.13126 \\
        Geom. & \phantom{$m$}2.8635 & \phantom{$mm$}1.3197 & \phantom{$mm$}1.5625 & \phantom{$-$}-0.01876 \\
        Hyper. & \phantom{$m$}1.3482 & \phantom{$mm$}1.3197 & \phantom{$mm$} 0.0000 & \phantom{$-$}-0.02844 \\
\hline
\end{tabular}
\vspace{0.3cm} \caption{Mean square error (and its components) for
four circle fits ($10^6 \times$values are shown). In this test
$n=10000$ points are placed (equally spaced) along a semicircle of
radius $R=1$ and the noise level is $\sigma = 0.05$.}
\label{Taballcir2}
\end{center}
\end{table}

\begin{figure}[htb]
    \centering
   \includegraphics[width=4in]{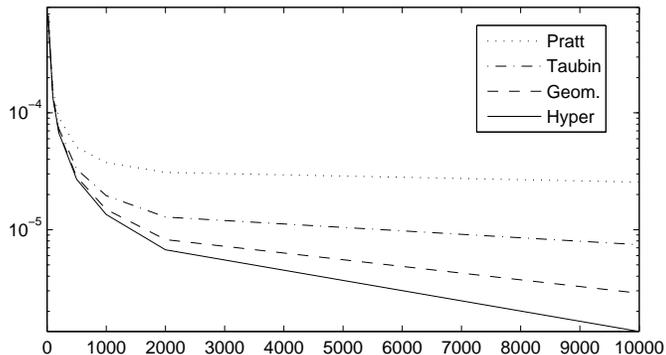}
    \caption{MSE for various circle fits (on the logarithmic scale)
    versus the sample size $n$ (from 10 to $10^4$).}
    \label{FigPlot}
\end{figure}

To highlight the superiority of the Hyper fit, we repeated our
experiment increasing the sample up to $n=10000$, see
Table~\ref{Taballcir2} and Figure~\ref{FigPlot}. We see that when
the number of points is high, the the Hyper fit becomes several
times more accurate than the geometric fit. Thus, our analysis
disproves the popular belief in the statistical community that there
is nothing better than minimizing the orthogonal distances.

Needless to say, the geometric fit involves iterative
approximations, which are computationally intensive and subject to
occasional divergence, while our Hyper fit is a fast non-iterative
procedure, which is 100\% reliable.

\medskip\noindent\textbf{Summary}.
All the known circle fits (geometric and algebraic) have the same
variance, to the leading order. The relative difference between them
can be traced to higher order terms in the expansion for the mean
square error. The second leading term in that expansion is the
essential bias, for which we have derived explicit expressions.
Circle fits with smaller essential bias perform better overall. This
explains a poor performance of the K{\aa}sa fit, a moderate
performance of the Pratt fit, and a good performance of the Taubin
and geometric fits (in this order). We showed that while there is a
natural lower bound on the variance to the leading order (the KCR
bound), there is no lower bound on the essential bias. In fact there
exists an algebraic fit with zero essential bias (the Hyper fit),
which outperforms the geometric fit in accuracy. We plan to perform
a similar analysis for ellipse fitting algorithms in the near
future.

The authors are grateful to the anonymous referees for many helpful
suggestions. N.C.\ was partially supported by National Science
Foundation, grant DMS-0652896.


\end{document}